# ON THE BIOPHYSICAL INTERPRETATION
# OF LETHAL DNA LESIONS INDUCED BY IONIZING RADIATION

Pavel Kundrát[1,*] and Robert D. Stewart[2]

[1]Institute of Physics, Academy of Sciences of the Czech Republic, Na Slovance 2, 18221 Praha 8, Czech Republic
[2]Purdue University, School of Health Sciences, 550 Stadium Mall Drive, West Lafayette, IN 47907-2051, USA

**Although DNA damage is widely viewed as a critical target for the induction of cell killing by ionizing radiation, the exact nature of DNA damage responsible for these effects is unknown. To address this issue, the probability of forming lethal damage by single proton tracks, derived from published survival data for Chinese hamster V79 cells irradiated by protons with energies from 0.57 to 5.01 MeV, has been compared to estimated yields of clustered DNA lesions and repair outcomes calculated with Monte Carlo models. The reported studies provide new information about the potential relationship between the induction and repair of clustered DNA damage and trends in the expected number of lethal events for protons with increasing linear energy transfer (LET). A good correlation was found between the number of lethal events in V79 cells and the induction of double-strand breaks (DSBs) consisting of 3 or more elementary DNA lesions. For the yields of other types of DNA damage, as well as point mutations formed through the misrepair of base damage and single-strand breaks, observed trends with increasing LET are not consistent with trends in the yields of lethal events. This observation suggests that the relative biological effectiveness (RBE) of protons of varying quality may be more closely related to the induction of complex DSBs rather than other forms of damage.**

## INTRODUCTION

Cell killing by ionizing radiation can be attributed to at least two distinct mechanisms: DNA damage that results in reproductive cell death and cell communication leading to the induction of apoptosis. Reproductive death is primarily attributed to lethal exchange-type chromosome aberrations, such as centric rings and dicentrics [1], which arise from the pairwise mis-rejoining of DNA double-strand breaks (DSBs). Differences in the yields and complexity of the initial DNA damage formed by radiations of diverse quality are often assumed to account for differences in biological effectiveness [2, 3]. Although over 95% of the initial DSBs formed by radiation are rejoined, complex DSBs may be intrinsically unrepairable and lethal. Alternatively, the slower rate of complex DSB rejoining [4] may increase the chance that pairs of DSBs interact to form lethal exchange-type aberrations. Mis-repaired or unrepairable single-strand breaks (SSBs) and sites of multiple damaged bases may also contribute to cell killing.

In the present work, the yields of single-track lethal events, derived from experimental survival data for V79 Chinese hamster cells after irradiation by protons at several LET (linear energy transfer) values, have been compared with estimated yields of different classes of complex DNA lesions calculated using Monte Carlo methods. These comparisons have been motivated by the effort to help elucidate some of the putative mechanisms of DNA damage-related cell killing and their dependence on radiation quality.

## METHODS

Monte Carlo simulations of different classes of damage to DNA have been performed using the fast Monte Carlo damage simulation algorithm (MCDS) [5]. The algorithm gives not only the yields of different damage classes but also their configurations. Repair outcomes for DNA damage configurations other than the DSB were computed using the Monte Carlo excision repair model (MCER) [6, 7] in combination with the MCDS algorithm. In addition to providing estimates of the number of point mutations formed through the misrepair of SSB and sites of multiple base damage, the MCDS/MCER simulations provide estimates of the number of DSBs formed through the aborted excision repair of complex SSBs (i.e., enzymatic DSBs).

The yields of lethal lesions induced by 0.57 to 5.01 MeV protons (LET values 7.7 – 37.8 keV/µm) in V79 cells have been derived from an analysis of survival data [8, 9] using the probabilistic two-stage model [10-12] and analysis method proposed in Ref. [11]. In this approach, the induction of DNA damage and any subsequent repair processes are treated collectively in the model through two test functions:

$$a(L)=a_0(1-\exp(-(a_1L)^{a_2})), \quad b(L)=b_0(1-\exp(-(b_1L)^{b_2})). \quad (1)$$

Here, $L$ denotes the LET of the protons and $a_0$, $a_1$, $a_2$, $b_0$, $b_1$ and $b_2$ are adjustable parameters. The $a(L)$ test function denotes the per track probability of inducing lethal damage through one-track mechanisms and $b(L)$ denotes the per track probability of multi-particle lethal damage [10, 11]. The expected number of one-track lethal lesions $Gy^{-1}$ $cell^{-1}$ is obtained by multiplying $a(L)$ by the average number of tracks $h$ passing through the cell nucleus per Gy [11]. In comparison with Ref. [11], where $a(L)$ was underestimated at low LET, an improved fit is used in the present work ($a_0$=0.041, $a_1$=0.050 μm/keV, $a_2$=2.58).

*Corresponding author: Pavel.Kundrat@fzu.cz

RESULTS AND DISCUSSION

Figure 1 compares the estimated yields of different classes of DNA damage (DSBs, SSBs, and other lesions) to the initial yield of one-track lethal events determined using the two-stage model and the linear-quadratic (LQ) survival model (reviewed in Ref. [13]). Also shown are the predicted numbers of enzymatic DSBs formed through the aborted excision repair of complex SSBs and the numbers of SSBs and base damages converted to point mutations through excision repair. The initial numbers of SSBs and sites of base damage (non-strand break lesions) exhibit a downward trend as the LET of the protons increases. However, the initial DSB yield shows an upward trend with increasing LET. For the range of proton energies considered, the predicted yield of prompt DSBs is 70 – 120 $Gy^{-1}$ $cell^{-1}$. The numbers of enzymatic DSBs are almost identical with the yields of mis-repaired lesions; for both lesion categories their yields increase with LET from approx. 40 to 70 $Gy^{-1}$ $cell^{-1}$. The present results are based on simulating DNA damage configurations and their repair outcomes in a total of $10^6$ cells for each proton energy (LET value); standard errors of the mean are smaller than the symbols. It has been assumed that the repair of all non-DSB lesions proceeds via the short-patch base excision repair (SP BER) pathway.

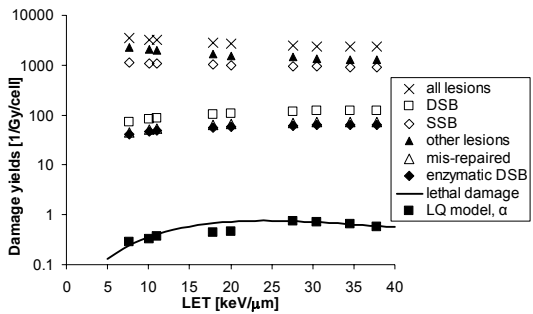

Figure 1: Comparison of one-track lethal damage to the initial yield of DSBs, SSBs and other DNA damage clusters for protons with energies from 0.57 to 5.01 MeV. Estimated numbers of mis-repaired non-DSB clusters converted to point mutations through the SP BER ("mis-repaired") and the yields of enzymatic DSBs following from the aborted repair of complex SSBs. Calculations according to the MCDS [5] and MCER models [6, 7]. Solid line denotes one-track lethal damage predicted with the two-stage model; for comparison, estimates of $\alpha$ from the LQ model reported in Refs. [8, 9] are shown, too.

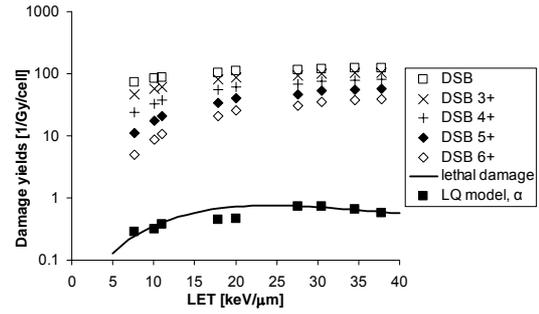

Figure 2: One-track lethal damage (as in Figure 1) compared to the yields of DSBs and subsets of DSBs of increasing complexity (DSB 3+, 4+, 5+, and 6+, composed of at least 3, 4, 5, or 6 elementary DNA lesions, respectively).

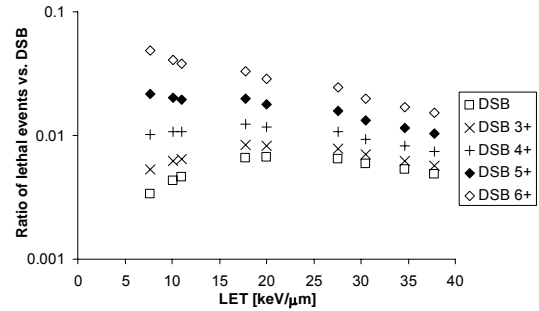

Figure 3: The ratio of lethal events among all DSBs and among DSBs of increasing complexity, neglecting the potential contribution of non-DSB lesions to cell killing.

The yields of lethal events arising in V79 cells through one-track mechanisms after proton irradiation range from 0.25 to 0.75 per Gy per cell. The yields of one-track lethal events increase with LET up to a maximum around 25 keV/μm and slowly decrease for higher LET. The comparison of one-track lethal events and damage yields in Figure 1 demonstrates that the vast majority of all classes of DNA damage, including DSBs, are repaired in a way that is non-lethal.

Complex DSBs are believed to pose significant challenges for cellular repair systems [2, 3]. To test the hypothesis that complex DSBs are the main damage responsible for cell killing, the yields of prompt DSBs simulated by the MCDS algorithm were sub-divided into DSBs of varying complexity, i.e. according to the number of elementary DNA lesions (strand breaks, base damage or abasic sites) per cluster. Figure 2 shows estimates of the yields of DSBs consisting of at least 3, 4, 5, and 6 elementary DNA lesions (denoted by DSB 3+, 4+, 5+, and 6+), respectively. Figure 3 shows the ratio of the expected number of lethal events per DSB, neglecting the cell killing effects of any mis-repaired lesions other than DSBs. The ratios of lethal events per highly complex DSBs (e.g. DSB 6+) exhibit decreasing trends with increasing LET, indicating that e.g. the DSB 6+ category of damage does not include all specific classes of damage involved in cell killing. The ratio of lethal events among all DSBs is largely dependent on LET, too. The least dependence on LET has been found for the ratios of lethal events per DSB 3+ and 4+, which suggests that this class of DSBs is a candidate for the damage most responsible for initiating cell death. The initial yields of DSB 3+ or 4+ are two orders of magnitude higher than the number of lethal events, i.e., only a subset of this class of DSB (about 0.7 – 1.2%) is converted to a lethal form of damage in V79 cells.

In the present work, the potential contribution of clusters other than DSBs to radiation-induced cell killing has been neglected. Most if not all enzymatic DSBs, formed through the aborted repair of complex SSBs, will be subsequently rejoined by non-homologous end-joining (NHEJ) or homologous recombination (HR) processes. Nevertheless, given that the yields of mis-repaired clusters and enzymatic DSBs are comparable to the numbers of prompt DSBs (Figure 1), this issue needs to be examined carefully in the future. The yields of enzymatic DSBs and especially of mis-repaired clusters is expected to be even higher if the long-patch base excision repair (LP BER) pathway is involved in the repair of clustered DNA lesions (Figure 4 and the discussion in Ref. [7]).

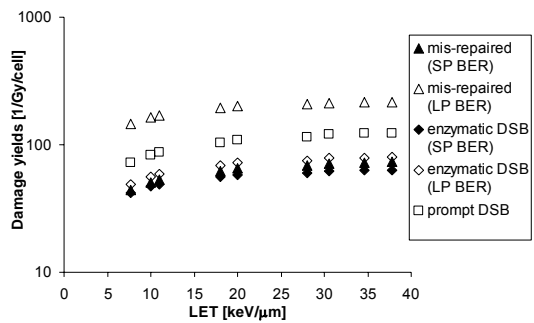

Figure 4: Yields of mis-repaired SSBs and non-strand break lesions (i.e., point mutations), as estimated for the short- and long-patch base excision repair pathways (SP, LP BER), respectively. Yields of prompt DSBs shown for comparison.

CONCLUSION

The reported results support the hypothesis that clustered DNA lesions formed by ionizing radiation play an important role in reproductive cell death. For V79 cells irradiated by protons with LET less than 40 keV/μm, a good correlation was found between the yield of one-track lethal events and the induction of DSB composed of 3 or more lesions. Further studies are necessary to elucidate the potential impact on cell killing of point mutations and enzymatic DSBs.

**Acknowledgment:** This work was supported in part by the grant "Modelling of radiobiological mechanism of protons and light ions in cells and tissues" (Czech Science Foundation, GACR 202/05/2728).